\documentclass[article,oneside,twocolumn,amsmath, prd,amssymb,showpacs,nofootinbib,superscriptaddress]{revtex4-2}

\usepackage{parskip}
\usepackage{amssymb}
\usepackage{amsfonts}
\usepackage{hhline}
\usepackage{amsmath}
\usepackage{mathtools}
\usepackage{xcolor}
\usepackage{xspace}
\usepackage{multirow,tabularx}
\usepackage{siunitx}
\usepackage{multirow}
\usepackage{graphicx}
\usepackage{xstring}
\usepackage{etoolbox}
\usepackage{notoccite}
\usepackage{natbib}
\usepackage{lineno}   
\usepackage{tikz}
\usepackage{listings}
\usepackage{braket}
\usepackage{slashed}
\usepackage{hyperref}
\usepackage[capitalize]{cleveref} 
\allowdisplaybreaks
\newcommand{\be}{\begin{equation}}
\newcommand{\ee}{\end{equation}}
\newcommand{\ba}{\begin{eqnarray}}
\newcommand{\ea}{\end{eqnarray}}

\hypersetup{colorlinks=true,
            breaklinks=true,
            pdfstartview=Fit,
            linkcolor=blue,
            citecolor=blue,
            urlcolor=blue}

\begin{document}

\title{Interact or Twist: Cosmological Correlators from Field Redefinitions Revisited}

\author{Dong-Gang Wang}
\affiliation{Department of Physics, The Hong Kong University of Science and Technology, Clear Water Bay, Kowloon, Hong Kong, P.R. China} 
\affiliation{The HKUST Jockey Club Institute for Advanced Study,
The Hong Kong University of Science and Technology,
Clear Water Bay, Kowloon, Hong Kong, P.R. China}
\affiliation{Centre for Theoretical Cosmology, Department of Applied Mathematics and Theoretical Physics,
University of Cambridge,
Wilberforce Road, Cambridge, CB3 0WA, U.K.}

\author{Xiangwei Wang}
\author{Yi Wang}
\author{Wenqi Yu}
\affiliation{Department of Physics, The Hong Kong University of Science and Technology, Clear Water Bay, Kowloon, Hong Kong, P.R. China} 
\affiliation{The HKUST Jockey Club Institute for Advanced Study,
The Hong Kong University of Science and Technology,
Clear Water Bay, Kowloon, Hong Kong, P.R. China}

\begin{abstract}

In cosmology, correlation functions on a late-time boundary can arise from both field redefinitions and bulk interactions, which are usually believed to generate distinct results. In this letter, we propose a counterexample showcasing that correlators from local field redefinitions can be identical to the ones from bulk interactions. In particular, we consider a two-field model in de Sitter space, where the field space gets twisted by field redefinitions to yield a nontrivial reheating surface. We then exploit conformal symmetry to compute the three-point function, and show that the result takes the form of contact correlators with a total-energy singularity. Our finding suggests that in the effective field theory, a class of lower-dimensional operators, which were overlooked previously, may lead to nontrivial signals in cosmological correlators. As an illustration, we apply our result to cosmic inflation and derive a possibly leading signature of the Higgs in the primordial bispectrum.

\end{abstract}

\maketitle

\section{Introduction}
\label{sec:introduction}

The studies on cosmological correlation functions provide an unsurpassed opportunity for deciphering the beginning of our Universe. 
Remarkably, these spatial correlations of the large-scale matter distribution, which depict the statistics of primordial fluctuations from cosmic inflation, can be tested in cosmological observations.
In recent years, by incorporating knowledge from conformal field theory (CFT), scattering amplitudes and the bootstrap approach, significant progress has been made in the precise understanding of the cosmological correlators \cite{Baumann:2022jpr}. It has been increasingly appreciated that
there are rich structures in these observables which encode the physics of inflation. 
One particularly interesting incarnation of this idea is the so-called ``cosmological collider physics" \cite{Chen:2009zp, Baumann:2011nk, Noumi:2012vr, Arkani-Hamed:2015bza}. 
In analogy with the terrestrial particle colliders,  this proposal suggests that massive states during inflation lead to characteristic signatures in the shapes of primordial non-Gaussianities (PNG)
\cite{Chen:2009we,Chen:2016nrs,Chen:2016nrs1,Meerburg:2016zdz,Chen:2016uwp,Chen:2016hrz,Kumar:2017ecc, Hook:2019vcn,An:2017hlx,Wu:2018lmx,Lu:2019tjj,Hook:2019zxa,Kumar:2019ebj,Wang:2019gbi,Lu:2021gso,Wang:2021qez,Tong:2021wai,Reece:2022soh,Pimentel:2022fsc,Jazayeri:2022kjy,Chen:2022vzh,Qin:2022lva,Wang:2022eop,Jazayeri:2023xcj,Pajer:2024ckd,Aoki:2024uyi,Pimentel:2025rds,Wang:2025qww,Qin:2025xct}, which have already been tested in the latest observational data \cite{Cabass:2024wob,Sohn:2024xzd}.

Unlike the S-matrix in flat spacetime, cosmological observables are defined on a spacelike boundary, normally specified as the end of inflation, or the reheating surface. In general, the equal-time correlators are not only outcomes of field interactions during inflation, but also affected by field redefinitions and boundary terms.
These two types of contributions are expected to yield different structures in the final correlators. 
For field redefinitions, within single field scenarios, correlators are highly constrained by scale invariance and locality \cite{Maldacena:2002vr,Pajer:2020wxk}; while for multi-field inflation, they were extensively investigated using the $\delta N$ formalism \cite{Starobinsky:1985ibc,Sasaki:1995aw,Lyth:2005fi}.
As only massless scalars are involved, these early studies normally led to the local shape in cosmological correlators.
Consequently, in recent advances on the theory frontier, most of the analysis focused on bulk interactions,
while the effects of field redefinitions with additional degrees of freedom have been relatively overlooked.

On the phenomenology side, the form of coupling between the inflaton and charged particles is restricted by charge conservation. As a result, in the cosmological collider signal of the Standard Model (SM) \cite{Chen:2016uwp,Chen:2016hrz,Kumar:2017ecc,Hook:2019vcn}, the dimension-six coupling $(\partial\phi)^2h^2$ between the inflaton $\phi$ and the Higgs $h$ was considered (where the real part of the Higgs field has a $Z_2$ symmetry), and the resulting loop diagrams are challenging to compute. However, as we discussed in the last paragraph, taking field redefinitions into consideration, one may wonder: (i) Can lower-dimensional (dimension-five) operators play a role? (ii) Can the loop diagrams be calculated more neatly, considering the simple nature of field redefinition?

In this Letter, based on the above observations, we shall re-examine field redefinitions involving more than one degrees of freedom and highlight their effects on cosmological correlators which were overlooked. In particular, we illustrate our finding through one simple example -- the mixing between the inflaton and a Higgs field through a dimension-five operator $h\partial_\mu h \partial^\mu \phi$.
Certainly this operator is more significant from the EFT power counting, but it can be removed by the following field redefinition
\be \label{phiss}
\phi \rightarrow \tilde\phi +  h^2 ,
\ee
after which $\tilde{\phi}$ and $h$ decouple.  
Surprisingly, the correlator from \eqref{phiss} becomes identical to the result of certain contact interactions in the bulk.
We perform the analysis in two parts. The first part establishes the connection with CFT in momentum space. Through a toy model, we explicitly show that field redefinitions with composite operators on the boundary lead to three-point conformal correlators in the Euclidean space, which
take the form of triple-$K$ integrals.
In the second part, we turn to a more realistic setup for cosmic inflation with a Higgs field.
Using field interactions, we identify non-trivial cosmological collider signals of the Higgs in the scalar bispectrum, as long as its mass is much lower than the Hubble scale during inflation. We also point out that our field redefinition procedure can be considered as a particle-theoretical realization of a non-trivial reheating surface in field space.

\section{Field redefinition \& CFT}
\label{sec:the model}
Let's start with a toy model of quantum field theory in a fixed 4D de Sitter (dS) spacetime and show how field redefinitions lead to conformal correlators on the boundary. We work in the flat slicing with conformal time, with the scale factor $a(\eta)=-1/H\eta$.
We consider a two-field system including a massless scalar field $\phi$ with a shift symmetry, and also an additional field $\sigma$ which enjoys a $\mathbf{Z}_2$ symmetry. 
The Lagrangian is written as 
\begin{equation} \label{theory}
    \mathcal{L}=-\frac12(\partial\phi)^2-\frac12\left(1+\frac{\sigma^2}{\Lambda^2}\right)(\partial \sigma)^2-\frac{1}{2}m^2 \sigma^2+\frac{1}{\Lambda}\sigma\partial_\mu \sigma \partial^\mu\phi  .
\end{equation}
The coupling between $\phi$ and $\sigma$ is given by a dimension-five operator that respects both symmetries. 
This interaction can be reduced to a boundary term by doing integration by parts (see Appendix \ref{app:ibp}).
Instead, here we perform the following field redefinition 
\be \label{phis2}
\phi = \tilde{\phi} + \frac{1}{2\Lambda}\sigma^2 ,
\ee
after which both this mixing and the quartic self-interaction of $\sigma$  are fully removed. We are left with a free theory for massless scalar $\tilde\phi$ and $\sigma$.
In this work, we are interested in $\sigma$ being a light scalar in the complimentary series ($0\leq m<3H/2$). Then on the late-time boundary of dS, its leading fall-off is given by $\lim_{\eta_0\rightarrow0^-}\sigma (\eta_0,{\bf x})\rightarrow \eta_0^{3/2-\nu} \mathcal{O}_{\Delta_\sigma}({\bf x})$ with $\nu\equiv \sqrt{9/4-m^2/H^2}$. Here $\mathcal{O}_{\Delta_\sigma}$ behaves as a conformal operator with a scaling dimension $\Delta_\sigma = \frac{3}{2}-\nu$.
Now if we compute the three-point function of the original massless scalar field $\phi$, we find that it is completely determined by the field redefinition \eqref{phis2}. The result takes a simple form in position space
\begin{align} \label{cft3pt}
\langle \phi({\bf x}_1) \phi({\bf x}_2) \phi({\bf x}_3)  \rangle =& \frac{1}{8\Lambda^3} \langle \sigma({\bf x}_1)^2 \sigma({\bf x}_2)^2 \sigma({\bf x}_3)^2 \rangle \\
=& \frac{ C_{\sigma}^3\eta_0^{6\Delta_\sigma}/{8\Lambda^3}}{x_{12}^{2\Delta_\sigma}x_{23}^{2\Delta_\sigma}x_{31}^{2\Delta_\sigma}} .\nonumber
\end{align}
with $ x_{ij}=|{\bf x_{i}}-{\bf x_j}|$.
In the second line we have used Wick contractions and $\langle  \sigma({\bf x}) \sigma({\bf x}')\rangle = C_{\sigma}\eta_0^{2\Delta_\sigma} /|{\bf x}-{\bf x}'|^{2\Delta_\sigma}$ with $C_\sigma$ being a constant determined by the mass $m$.

It is interesting to notice that this boundary correlator of a massless scalar takes the same form as the three-point function $\langle \mathcal{O}_\Delta({\bf x}_1) \mathcal{O}_\Delta({\bf x}_2) \mathcal{O}_\Delta({\bf x}_3) \rangle_{\rm CFT}$ in a 3d Euclidean CFT with $\Delta=2\Delta_\sigma$. 
The above relation can be explained as follows: for a free $\sigma$ field in complementary series, the composite operator $\sigma({\bf x})^2$ on the boundary becomes $\sigma^2 \rightarrow \eta_0^{2\Delta_\sigma} \mathcal{O}_\Delta + ...$, and the leading fall-off corresponds to a CFT operator with scaling dimension $\Delta = 2\Delta_\sigma$. This observation suggests that the three-point correlator \eqref{cft3pt} satisfies all conformal symmetries, which will significantly simplify the subsequent analysis. 

\begin{figure}[t]
    \centering
    \includegraphics[width=0.95\linewidth]{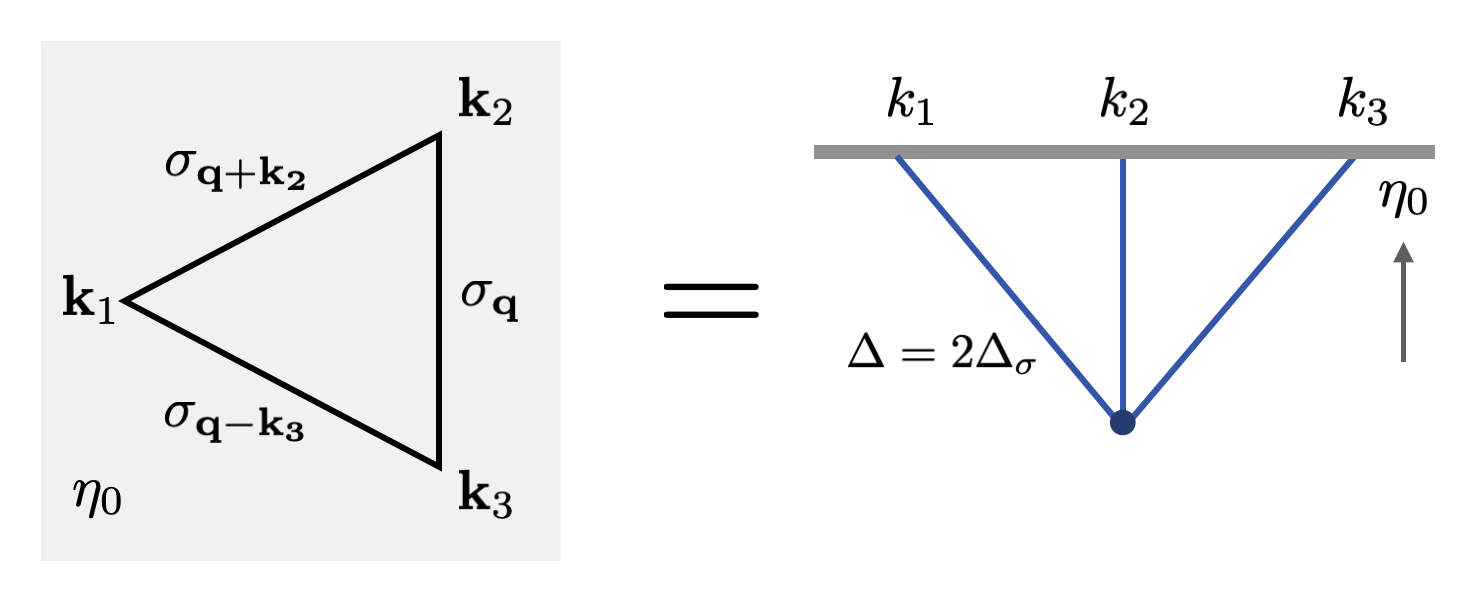}
    \caption{The boundary ``one-loop" triangle graph from field redefinition \eqref{bispectrumfieldredefi} can be re-expressed as a contact interaction of fields with a different mass in the bulk \eqref{tripleK}.}
    \label{field rede one loop}
\end{figure}

Next, to compute the inflaton bispectrum, we may directly work in momentum space. Then the field redefinition contribution corresponds to a ``one-loop" triangle graph on the future boundary (see Figure \ref{field rede one loop})
\begin{align}\label{bispectrumfieldredefi}
\langle \phi_{{\bf k}_1}\phi_{{\bf k}_2}\phi_{{\bf k}_3} \rangle'= 
& \left(-\frac{\eta_0}{2}\right)^{9-6\nu}\frac{H^6\Gamma(\nu)^6}{\pi^3\Lambda^3} I_\nu({k}_1,{k}_2,{k}_3) ,
\\
{\rm with}~~I_\nu\equiv \int&\frac{d^3\mathbf{q}}{(2\pi)^3} \frac{1}{
|\mathbf{q}|^{2\nu}|\mathbf{q}+\mathbf{k}_2|^{2\nu}|\mathbf{q}-\mathbf{k}_3|^{2\nu}} , \label{Inu}
\end{align}
where $'$ means that the momentum conservation $\delta$-function has been stripped off.
This is a complicated momentum integral for general values of $\nu\in (0,3/2]$.
\footnote{This momentum integral also appears when one evaluates triangle loops with massive internal lines using the Mellin space approach \cite{Qin:2023bjk}. The simplification here is that we do not have bulk time integrals when performing field redefinitions.}
To circumvent this challenge, we can perform the Fourier transformation of the three-point conformal correlator \eqref{cft3pt}.  The advantage of this approach is that the conformal symmetries become manifest and we can exploit the systematic studies on CFT correlators in momentum space \cite{Bzowski:2013sza,Coriano:2013jba,Bzowski:2015pba}. 
From there, we know that $I_\nu(k_1,k_2,k_2)$ satisfies the conformal Ward identities of dilatation and special conformal transformations, which lead to two second-order partial differential equations. The solution can be expressed as a linear combination of the Appell $F_4$ functions, which are generalized hypergeometric functions with two variables. We leave the full expression in \eqref{F4}.

The solutions of conformal Ward identities in momentum space also have an integral representation 
\begin{small}\begin{align} \label{tripleK}
  I_\nu =& \frac{\pi^{-\frac32}2^{-\frac12}}{\Gamma\left(\frac{3\Delta-3}{2}\right)\Gamma\left(\frac{3-\Delta}{2}\right)^3}(k_1k_2k_3)^{\Delta-\frac32}\int_{-\infty}^0 d\eta 
(-\eta)^{1/2}  \nonumber\\
&\times K_{\Delta-3/2}(-k_1\eta) K_{\Delta-3/2}(-k_2\eta)K_{\Delta-3/2}(-k_3\eta) ,
\end{align}
\end{small}where $ K_{\Delta-3/2}$ is the Bessel $K$ function.
This is known as the triple-$K$ integral, which takes the form of the contact interaction of three massive scalars with $\Delta=2\Delta_\sigma$ in the bulk. 
As an illustration, let's compute the explicit results of this integral for two specific cases of $\sigma$ field: massless ($\Delta_\sigma =0$) and conformally coupled ($\Delta_\sigma =1$) scalars
\begin{align}
I_{\frac32}=& \frac{\left(\frac43-\gamma_E-\log{k_T/\mu}\right)\mathop{\sum}\limits_{i=1}^3k_i^3+\mathop{\sum}\limits_{i\neq j}k_i^2k_j-k_1k_2k_3}{2\pi^2k_1^3k_2^3k_3^3}~,\nonumber\\
I_{\frac12}=& \frac1{2\pi^2}\left(-\gamma_\mathrm{E}-\log{{k_T}/{\mu}}\right) ,\nonumber
\end{align}
where we have renormalized the infrared divergences here by adding counterterms with a scale $\mu$.
For general masses, we always find a logarithmic pole in the $k_T=k_1+k_2+k_3\rightarrow0$ limit  
\be \label{kT}
\lim_{k_T\rightarrow0}I_\nu = -\frac{(k_1k_2k_3)^{\Delta-2}}{4\Gamma\left(\frac{3\Delta-3}{2}\right)\Gamma \left(\frac{3-\Delta}{2}\right)^3} 
\log{k_T}.
\ee
It is normally expected that, for cosmological correlators, this total energy singularity is generated by bulk interactions, as the residue of the pole corresponds to flat-space amplitudes \cite{Maldacena:2011nz,Raju:2012zr}, while field redefinitions can only affect parts of the correlators that are regular in the $k_T\rightarrow0$ limit. The computation here serves as a counterexample showing that this singularity can also be a consequence of  field redefinitions with composite operators.

In summary, we show that the momentum integrations of composite operators from a {\it boundary} field redefinition lead to the same result from a {\it bulk} contact interaction.  
From the boundary perspective of the cosmological bootstrap, for dS-invariant theories, once we know the scaling dimensions, the three-point functions are fully determined by conformal Ward identities \cite{Maldacena:2011nz,Arkani-Hamed:2015bza,Arkani-Hamed:2018kmz,Baumann:2020dch}. Thus, it may not be surprising that the correlators are in the form of triple-$K$ integrals.
However, one main novelty arising at the loop level is the shift of scaling dimensions. 
With interactions in the bulk, in general we expect $\Delta\rightarrow \Delta+\gamma$ with $\gamma>0$ being anomalous dimensions from loop corrections \cite{Marolf:2010zp,Arkani-Hamed:2015bza,Cohen:2024anu}. 
When field redefinitions are involved, the bulk field becomes a sum of 3d conformal operators on the boundary, each with its own scaling dimension.
For the three-point function in our consideration, the leading contribution arises at one-loop, where the scaling dimension simply inherits the one of the composite operator $\sigma({\bf x})^2$.
Thus, under the requirement of dS isometries, 
once the scaling dimensions are fixed, field redefinitions and bulk interactions are forced to give the same boundary three-point correlators.

More generally, we may consider the following field redefinition involving primary CFT operators with scaling dimension $\Delta_n$ on the Euclidean boundary
\be
\phi = \tilde\phi + \sum_n\eta_0^{\Delta_n} \mathcal{O}_{\Delta_n}~.
\ee
Then regardless of the bulk theory, the above computation of the $\phi$ bispectrum would still apply, and the results in general can be expressed as the ones from contact interactions of bulk fields with different masses.\footnote{For a particular example, we may consider a higher-order bulk field redefinition like $\phi = \tilde{\phi} + \sigma^2 + \sigma^3 $, which can contribute to a correlator of three composite operators with $\Delta_1=2\Delta_\sigma$ and $\Delta_2=\Delta_3=3\Delta_\sigma$:
\begin{align}
\langle \phi({\bf x}_1) \phi({\bf x}_2) \phi({\bf x}_3)  \rangle \supset & \langle \sigma({\bf x}_1)^2 \sigma({\bf x}_2)^3 \sigma({\bf x}_3)^3 \rangle \nonumber \\ \sim & \langle \mathcal{O}_{\Delta_1}  \mathcal{O}_{\Delta_2} \mathcal{O}_{\Delta_3} \rangle_{\rm CFT} . \nonumber
\end{align}
}
Strictly speaking, at the late-time limit $\eta_0\rightarrow0$, this conformal three-point correlator is non-vanishing only if $\Delta=0$. For non-zero values of $\Delta$, the bispectrum is suppressed in terms of a power-law of $\eta_0$ as shown in \eqref{bispectrumfieldredefi}. 
However, for a non-zero late-time cut-off and a small $\Delta$, the size of the signal may still be significant.
We leave more discussion in the phenomenology section.

\section{Leading signature of a Higgs}

Now we move to a more realistic scenario of cosmic inflation with the inflaton and a Higgs field. 

We treat the Higgs mass squared as a free parameter, since it receives corrections from Higgs self-coupling, coupling to the inflaton  and coupling $h^2R$ between the Higgs and the Ricci scalar. As a result, during inflation, the Higgs may: (i) have vanishing VEV, where the discussion is parallel to the previous section, so we will not repeat it here; or (ii) have nonzero VEV $\bar h\neq 0$ and the electroweak symmetry is spontaneously broken. 

In this section we will focus on case (ii), and denote the fluctuations around this VEV as $h$. The Higgs sector Lagrangian is given by $ \mathcal{L}_{h}=-\frac12(\partial h)^2+\mu^2 h^2-\frac{1}{4}\lambda(h^4+4\bar{h}h^3)$
with $\mu^2<0$ and $\bar h^2 = -\mu^2/\lambda$. 
In the spatially flat gauge and the decoupling limit where gravitational interactions are suppressed by slow-roll parameters, the leading interaction in this two-field system is given by the mixing between  $h$ and the inflaton fluctuation $\phi$ 
\be \label{Lmix}
  \mathcal{L}_{\rm mix.}= \frac{\bar{h}}{\Lambda}\partial_\mu h\partial^\mu\phi
  +\frac{1}{\Lambda}h\partial_\mu h\partial^\mu\phi .
\ee

This two-field system is similar to our toy model in the previous section. One difference is that $\mathcal{L}_h$ contains self-interactions of the $h$ field, which may lead to anomalous dimensions of the composite operators $h({\bf x})^n$ on the boundary. 
Another difference is the kinetic mixing caused by the non-zero $\bar h$. This can be removed together with the dimension-five operator by performing the following field redefinition
\be \label{phih2}
\phi = \tilde{\phi} + \frac{\bar h}{\Lambda} h + \frac{1}{2\Lambda}h^2 .
\ee
After this, the two scalars $\tilde\phi$ and $h$ become decoupled. 
Meanwhile, the kinetic term of $h$ acquires corrections $ (h+\bar{h})^2(\partial h)^2/{2\Lambda^2}$. Then for the canonically normalized Higgs field,
its mass becomes $ {m}^2 = -2\mu^2 / (1-\bar{h}^2/\Lambda^2)$. 

\begin{figure}
    \centering
    \includegraphics[width=0.82\linewidth]{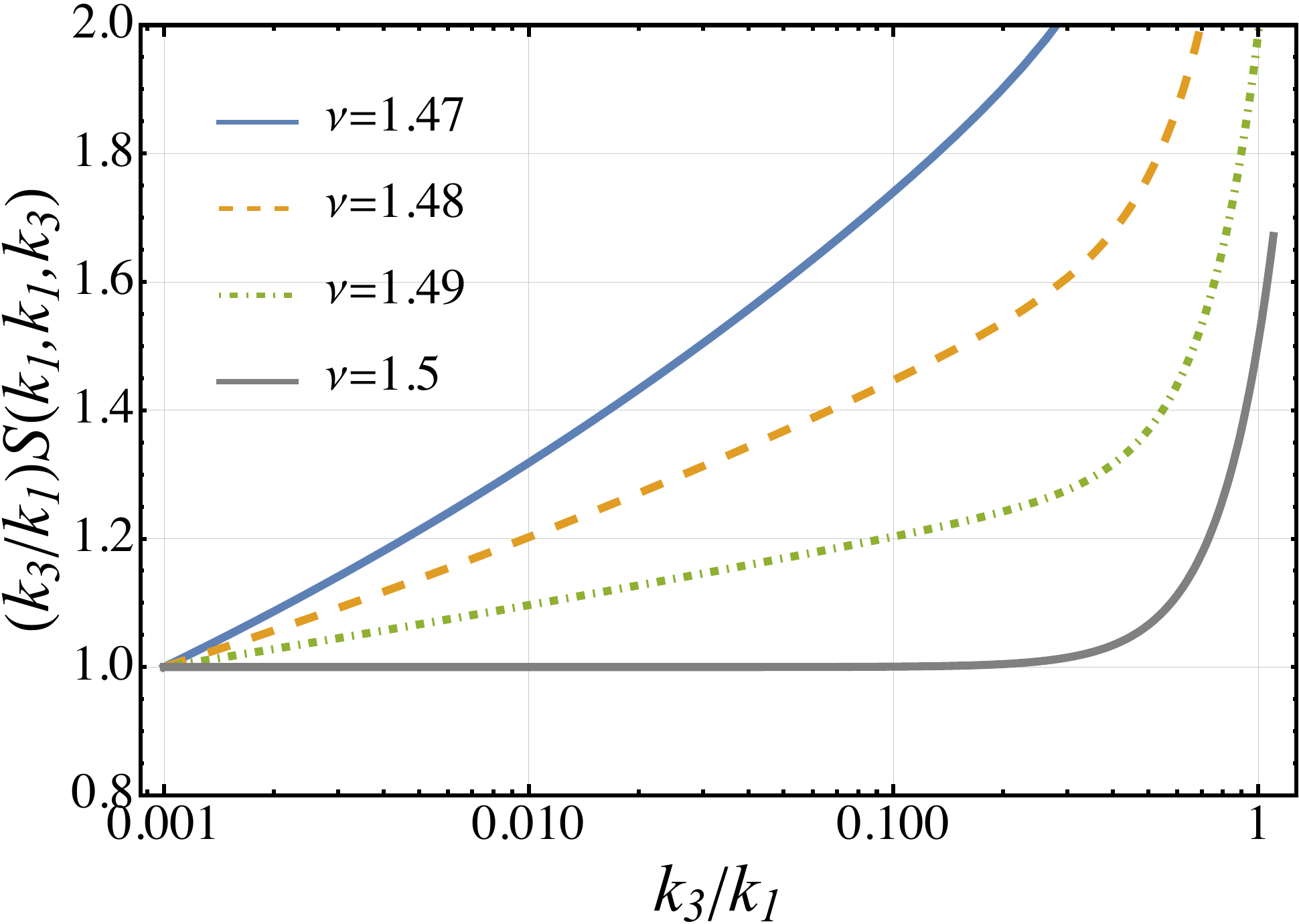}
    \caption{Bispectrum shape functions with different masses in the isosceles-triangle configuration with $k_1=k_2$. } 
    \label{loopshape}
\end{figure}

Next, for the primordial curvature perturbation $\zeta =\phi/(\sqrt{2\epsilon}M_{\rm pl})$ in this scenario, we compute its bispectrum as
$\langle \zeta_{{\bf k}_1}\zeta_{{\bf k}_2}\zeta_{{\bf k}_3} \rangle'\equiv f_{\rm NL} P_\zeta^2 S(k_1,k_2,k_3)/k_1^2k_2^2k_3^2 $, where $f_{\rm NL}$ is the size parameter, $S$ is the dimensionless shape function, and $P_\zeta\equiv H^2/(8\pi^2M_{\rm pl}^2\epsilon)$ is the primordial power spectrum.
We work under the assumption that the Higgs self-interactions are highly suppressed ($\lambda\ll 1$ and $\bar h\ll \Lambda$), such that the dominant contribution comes from the field redefinition \eqref{phih2}. 
Then the bispectrum has two components
\begin{equation}
\label{bispectrumfieldredefissb}
\braket{\zeta_1\zeta_2\zeta_3}=
\frac{\braket{h_1^2h_2^2h_3^2}}{(2\sqrt{2\epsilon}\Lambda M_{\rm pl})^3}
+\frac{\bar{h}^2}{2}\frac{\sum_{\text{perms}}\braket{h_1^2h_2h_3}}{(\sqrt{2\epsilon}\Lambda M_{\rm pl})^3}~,
\end{equation}
where we have used the shorthand notations $\zeta_i=\zeta({\bf x_i})$ and $h_i=h({\bf x_i})$.
The first term is basically what we have computed in \eqref{cft3pt}, which leads to a shape function $
        S_{\rm I}\propto
    k_1^2k_2^2k_3^2I_\nu(k_1,k_2,k_3)
$.
For the massless case ($\nu=3/2$), this becomes the same as the conformal non-Gaussianity that arises from the inflaton self-interaction $\phi^3$ \cite{Pajer:2016ieg}.
For general masses, its analytical expression is complicated, as we have seen in the above analysis and Appendix \ref{app:appell}.
We show several examples with light masses in Fig. \ref{loopshape}.
It is informative to look at its squeezed limit $k_3\ll k_1\simeq k_2$, where the shape function becomes $S^{\rm I}(k_1,k_2,k_3)\propto ({k_3}/{k_1})^{5-4\nu}$ for $3/4<\nu <3/2$. This can be seen as the scaling of two composite operators $h^2$.
The second term in \eqref{bispectrumfieldredefissb} is the standard result expected from a local field redefinition.
In Fourier space, it simply becomes
\begin{align}
    &\frac{\bar{h}^2}{2\Lambda^3} \langle h_{\bf{k}_2} h_{-\bf{k}_2} \rangle \langle h_{\bf{k}_2} h_{-\bf{k}_2} \rangle
    +2 \text{ perms}\\
    =&\frac{\bar{h}^2}{\Lambda^3}\frac{2 H^4(-\eta_0/2)^{6-4{\nu}}}{\pi^2\left(1-{\bar{h}^2}/{\Lambda^2}\right)^2}\Gamma({\nu})^4k_2^{-2{\nu}}k_3^{-2{\nu}}+2 \text{ perms}. \nonumber
\end{align}
Unlike \eqref{bispectrumfieldredefi}, this correlator contains only one composite operator from field redefinition, and thus there is no momentum integral involved.
In the massless limit $\nu\rightarrow 3/2$, we get back to the well-known local non-Gaussianity.
For general masses, it becomes
\begin{equation}
    S_{\rm II}=\frac{\left({k_2}/{k_*}\right)^{3-2\bar{\nu}}\left({k_3}/{k_*}\right)^{3-2\bar{\nu}}k_1^3}{3k_1k_2k_3} +2 \text{ perms}~,
\end{equation}
where $k_*$ is the pivot scale for observations.
In the squeezed limit, we find $\lim_{k_3\rightarrow 0}S_{\rm II}\propto ({k_3}/{k_1})^{2-2\nu}$, which inherits the scaling of one composite operator $h^2$.

Now we take a look at the size of PNG. For a massive $h$ field, both terms in \eqref{bispectrumfieldredefissb} are exponentially suppressed at the late-time limit $\eta_0\rightarrow 0$: the first with $(-k_*\eta_0)^{9-6\nu}$ and the second with $(-k_*\eta_0)^{6-4\nu}$. As the pivot scale is normally chosen for CMB modes, we have $\log(-k_*\eta_0)\simeq 50-60$.
Thus only when the Higgs field is light, these signals may become detectable and the shape function there resembles the local one.
For $m\ll H$, the size parameter for the first term in \eqref{bispectrumfieldredefissb} becomes 
\begin{equation}
    f_{\rm NL}^{\rm I}\simeq\frac{9\pi^2  \sqrt{\epsilon}M_{\rm pl}H^4}{512 \Lambda^3 m^2}(-k_* \eta_0)^{\frac{2m^2}{H^2}} .
\end{equation}
For a benchmark choice of parameters: $m/H$ is $ \mathcal{O}(0.1)$ and $\Lambda/H$ is $\mathcal{O}(10)$, $f_{\rm NL}^{\rm I}\sim 1$ can be easily achieved. 
Similarly the second term leads to 
$f_\mathrm{NL}^{\rm II}\simeq\frac{3\pi{\sqrt{\epsilon }M_{\mathrm{pl}}\bar{h}^2}}{\Lambda^3(1-{\bar{h}^2}/{\Lambda^2})^2}(-{k_*\eta_0}/{2})^{{4m^2}/{3H^2}}$,
which has an extra dependence on the VEV of the Higgs field.

\section{Field redefinition \& Reheating surface}

From the perspective of multi-field inflation, the above model corresponds to a simple scenario that the inflaton trajectory is along a straight line in a flat 2d field space. Normally we would expect that this just reduces to single field models. Then how come we still find nontrivial PNG from field redefinitions?

The interpretation is as follows. In multi-dimensional field spaces, a field redefinition corresponds to a coordinate transformation, such as $(\phi,h)$ to $(\tilde\phi,h)$ in the above example. Then although we may perform computations in the new field basis, the inflaton field remains to be the original one and is responsible for  generating curvature perturbations.
For field redefinitions involving multiple fields, their effect is to {\it twist} the field space, which also deforms the the end of inflation. As a result, in the redefined theory, the generation of PNG is mainly due to the nontrivial reheating surface in the field space \cite{Sasaki:2008uc,Naruko:2008sq,Huang:2009vk}. There the extra field fluctuations affect the duration of inflation and thus in the end convert to the final curvature perturbations.
This conversion mechanism differs from the conventional one in multi-field inflation where the turning trajectory generates a quadratic interaction $\dot\phi\sigma$ during inflation between the inflaton fluctuations and the isocurvature modes \cite{Gordon:2000hv,Achucarro:2016fby,Wang:2022eop}.

\begin{figure}[t]
    \centering
    \includegraphics[width=0.6\linewidth]{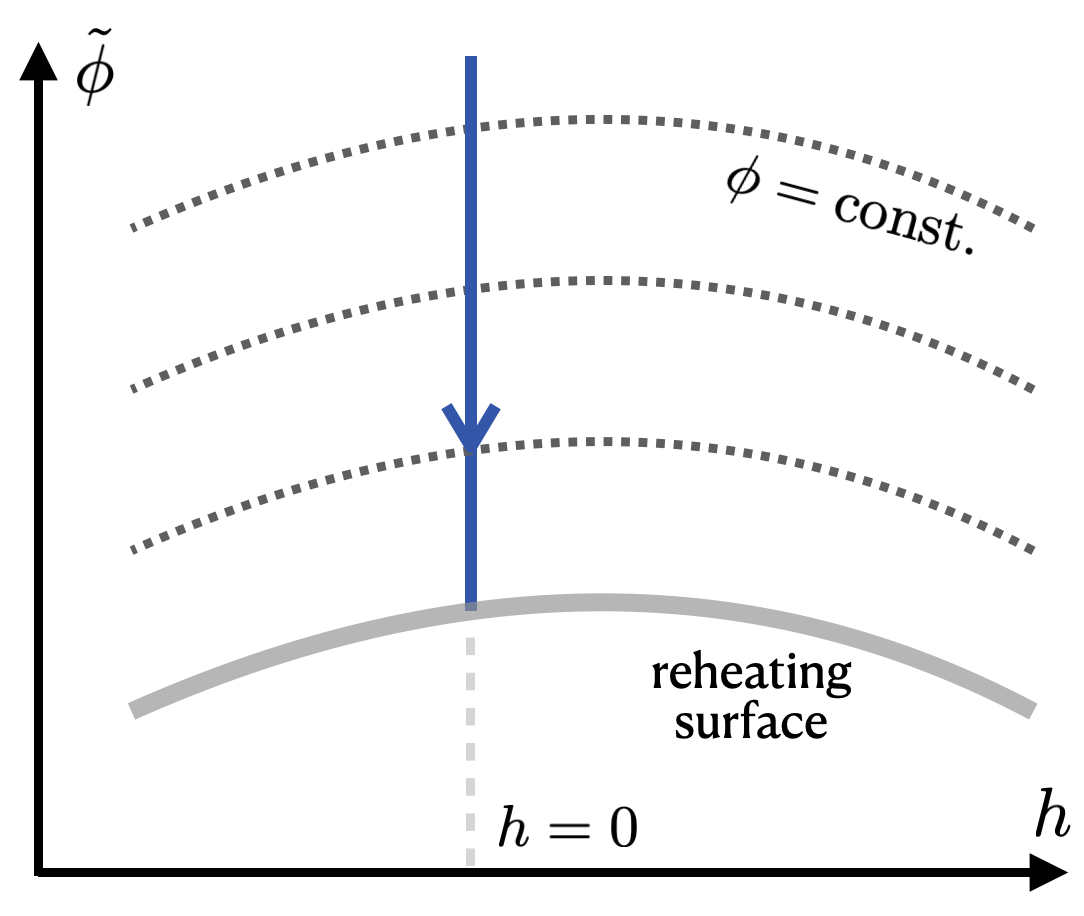}
    \caption{Field redefinitions twist the field space and lead to a nontrivial reheating surface in the new $(\tilde\phi,h)$ basis. For our example, the dotted lines are constant $\phi$ surfaces, while the inflaton trajectory (blue) is along $h=0$ in both field bases.}
    \label{fig:reh}
\end{figure}

In our model, the original Lagrangian corresponds to an unorthogonalized field basis due to the kinetic coupling \eqref{Lmix}.
After redefinition \eqref{phih2}, the $(\tilde\phi, h)$ basis becomes an orthogonal one as shown in Fig. \ref{fig:reh}, while the end of inflation is given by one of the twisted $\phi={\rm const.}$ surfaces.
Thus, although there is no field interaction during inflation, the nonlinearity of the conversion contributes to the three-point correlators in our computation. 
Note that one important requirement for nonvanishing bispectrum is that the reheating surface needs to be curved.
In particular, for $\bar h =0$, the inflaton trajectory is perpendicular to the reheating surface, we get the three-point function as the first term in \eqref{bispectrumfieldredefissb}, same as the one from bulk contact interactions.

\section{Discussions}\label{Discussion}

Cosmological correlators are field redefinition dependent. In previous studies, one common lore is that their contributions are simply of the local form which differs from the correlators from bulk interaction. Therefore, they have not been given much attention in the recent developments on the analytical structure of these cosmological observables. 
In this Letter, we propose a concrete example to challenge this oversimplified understanding. 

We start with a dimension-5 mixing between the inflaton and an additional scalar with $Z_2$ symmetry. 
Although this is a leading operator from the EFT power counting, it can be completely removed by a field redefinition and thus was neglected previously.
However, by performing the field redefinition, we identify that the resulting three-point function becomes identical to conformal correlators in the form of triple-$K$ integrals.
Therefore, contrary to the naive expectation, field redefinitions can lead to structures of cosmological correlators that are the same as those from bulk interactions. 

In practical terms, here we identify a possible leading signature of Higgs fields from field redefinitions in the context of cosmological collider physics. 
As a physical interpretation, we argue that the effect of these redefinitions is to twist the multi-dimensional field space, which leads to a nontrivial reheating surface in multi-field inflation.
This finding suggests us to re-think what are the leading mixing operators in the cosmological EFTs.
As a simple extension of our scalar example, we may take a look at massive spinning particles.
For fermions and spin-1 particles, the lowest-derivative parity-even couplings to the inflaton are $\partial_\mu\phi\bar{\psi}\gamma^\mu\psi$ and $\partial_\mu\phi\partial^\mu(A_\nu A^{\nu})$.\footnote{At the  dimension-5 order of the EFT, there are also parity-odd operators: the axial-vector current term $\partial_\mu\phi\bar{\psi}\gamma^5\gamma^\mu\psi$ and the Chern-Simons term ${\phi}\epsilon^{\mu\nu\rho\sigma}F_{\mu\nu}F_{\rho\sigma}$. Both of them cannot be removed by field redefinitions. }
The former vanishes under a phase rotation of the fermion $\psi\rightarrow\psi e^{i\phi}$, which does not affect the inflaton correlators; the latter can be analyzed by similar field redefinitions as the scalar example.
These effects are generally expected in more realistic setups of inflation and may lead to new types of signals in correlators.
We leave a more systematic investigation in future work.

In the end, this work may open up new lines of research in the analytical studies of cosmological correlators, especially for loop-level structures.
One technical simplification in the toy model \eqref{theory} is that the boundary momentum integral \eqref{Inu} can be re-expressed as a bulk-type triple-$K$ integral \eqref{tripleK}. 
This is mathematically nontrivial but can be seen as a natural consequence from the conformal symmetries. 
In this special example, the scaling dimensions are fixed by the composite operators in the field redefinition.
Also from the integration by parts analysis,
field interactions can be fully reduced to boundary terms and thus no bulk time evolution is involved (see Appendix \ref{app:ibp}).
For generic interactions, loop-level correlators contain both time and momentum integrations, which makes the analytical computations rather complicated. 
However, the constraints from dS isometries still apply, which forces the three-point correlators to be solutions of conformal Ward identities.
Roughly speaking, at loop-level these correlators might become a sum/integration of triple-$K$ integrals with some ``spectral density". 
This kind of spectral decomposition for two-point correlators at one-loop level has been noticed in \cite{Marolf:2010zp,Xianyu:2022jwk}, while our understanding about three-point correlators is still immature. An improved analysis will help us extend the cosmological bootstrap to loop level and identify the one-loop structure of three-point functions.

\section*{Acknowledgements}
We would like to thank Sebastian Cespedes, Andrew Cohen, Yanjiao Ma, Enrico Pajer, Guilherme Pimentel, Zhehan Qin, David Stefanyszyn, Xi Tong, Junqi Wang, Xingkai Zhang and Zhong-Zhi Xianyu for helpful discussions. This work was supported in part by the National Key R\&D Program of China (2021YFC2203100), and the RGC Research Fellow Grant RFS2425-6S02 from the RGC of Hong Kong.

\newpage

\begin{widetext}
\appendix

\section{Details of the Triangle-Loop Integral}
\label{app:appell}

In this Appendix, we present the full result of the momentum integral in \eqref{Inu} using the Fourier transformation of three-point CFT correlators, and investigate its analytical structure.
In terms of Appell $F_4$ functions, the full solution of conformal Ward identities in momentum space is given by the following linear combination \cite{Bzowski:2013sza,Coriano:2013jba}
    \begin{align} \label{F4}
        I_\nu &= \frac{k_1^{3-6\nu} }{(4\pi)^{\frac32}\Gamma(3\nu-3)\Gamma(\nu)^3}  \left(\frac{\pi}{\mathrm{sin}(2v-\frac12)\pi}\right)^2 \\
        &\times\left\{ F\left(3\nu-\frac32,\nu,2\nu-\frac12,2\nu-\frac12\middle|u^{2},w^{2}\right) -F\left(-\nu+\frac32,\nu , 2\nu-\frac12,-2\nu+\frac52\middle|u^{2},w^{2}\right)w^{4\nu-3}\right. \nonumber \\
       &\left.-F\left(-\nu+\frac32,\nu , -2\nu+\frac52,2\nu-\frac12\middle|u^{2},w^{2}\right)u^{4\nu-3} +F\left(-\nu+\frac32,-3\nu+3;-2\nu+\frac52,-2\nu+\frac52\middle|u^{2},w^{2}\right)\left(uw\right)^{4\nu-3} \right\},\nonumber
    \end{align}
with $u\equiv k_1/k_2$ and $w\equiv k_1/k_3$. The $F$ functions are defined by
\be
F(\alpha,\beta,\gamma,\gamma';u,w)\equiv \Gamma\left(\begin{matrix}
          \alpha, \beta \\ \gamma, \gamma'
       \end{matrix}\right)\mathcal{F}_4\left(  \alpha, \beta;\gamma, \gamma' |u^{2},w^{2}\right)~,
\ee
with $\mathcal{F}_4$ being the Appell $F_4$ function. There are four solutions of the conformal Ward identities for $I_\nu$, which correspond to the four terms of in \eqref{F4}. The coefficients are determined by the requirement of the absence of folded singularities at $k_1=k_2+k_3$.
This solution is equivalent to the triple-$K$ integral in \eqref{tripleK}, from which we can easily identify its singular behaviour at $k_T\rightarrow 0$ as shown in \eqref{kT}.
For the soft limit, it is more convenient to take a look at the isosceles-triangle configuration with $k_1=k_2\geq k_3$, where \eqref{F4} reduces to
\begin{align}
       I_{\nu}&=\frac{k_1^{3-6\nu} }{8\pi\Gamma(3\nu-3)\Gamma(\nu)^3}  \frac{2^{1-2\nu}}{\mathrm{sin}\left[2(v-\frac12)\pi\right]}\left[\Gamma\left(\begin{matrix}
        -\nu+\frac32,\nu,3\nu-\frac32\\2\nu-\frac12,\nu+\frac12
    \end{matrix}\right){}_3{F}_2\left(\begin{matrix}
        -\nu+\frac32,\nu,3\nu-\frac32\\2\nu-\frac12,\nu+\frac12
    \end{matrix}\middle|\left(\frac{k_3}{2k_1}\right)^2\right)\right.\\    &\left.~~~~~~~~~~~~~~~~~~~~~~~~~~~~~~~~~~~~~~~~~~~~~~~-\left(\frac{k_3}{2k_1}\right)^{3-4\nu}\Gamma\left(\begin{matrix}
        -\nu+\frac32,\nu,-3\nu+3\\-2\nu+\frac52,-\nu+2 
    \end{matrix}\right){}_3F_2\left(\begin{matrix}
        -\nu+\frac32,\nu,-3\nu+3\\-2\nu+\frac52,-\nu+2
    \end{matrix}\middle|\left(\frac{k_3}{2k_1}\right)^2\right)\right] . \nonumber
\end{align}
As the two ${}_3F_2$ hypergeometric functions approach unity with $k_3\rightarrow0$, the soft-limit scaling then demonstrates the following behaviour
\begin{equation}
\lim_{k_3\rightarrow0}I_\nu   \sim 
\begin{cases}
    k_3^{3-4\nu} & ~~ \frac{3}{4} <\nu <\frac{3}{2}  \\
    {\rm const.} &  ~~ 0<\nu<\frac{3}{4} 
\end{cases} .
\end{equation}
This suggests that in the squeezed limit the $\langle \phi_{{\bf k}_1}\phi_{{\bf k}_2}\phi_{{\bf k}_3} \rangle$ correlator cannot decrease faster than $k_3^0$.
We show the scalings of the bispectrum shapes for several light masses in
Figure \ref{loopshape}.
For half-integer $\mu$'s, like conformally coupled and massless scalars, we encounter logarithmic infrared divergences and renormalization is needed \cite{Bzowski:2015pba}.

\section{Boundary Term through IBP}
\label{app:ibp}

For each field redefinition, in general, we can find a boundary term that gives us the same result of cosmological correlators. 
In this Appendix, we shall redo the analysis for the toy model example \eqref{theory} by performing Integration-by-Part (IBP).
The cubic mixing there becomes
\begin{equation}\label{intLIBP1}
    \frac{1}{\Lambda}\sigma\partial_\mu\sigma\partial^\mu\phi=-\frac{1}{2\Lambda}\sigma^2\square{\phi} +\frac{1}{2\Lambda}\nabla_\mu(\sigma^2\partial^\mu \phi ) .
\end{equation}
On the right hand side, the first term vanishes at the one-loop level, as all the $\phi$ lines are external and on-shell, with $\square\phi=0$.
The second term is the boundary term that cannot be neglected. 
Since we are evaluating the correlators on a future boundary of dS with $\eta = \eta_0$, this term corresponds to an interacting Lagrangian
$\mathcal{L}_{\mathrm{int}}= ({1}/{2\Lambda a^2})\sigma^2\partial_\eta\phi \delta(\eta-\eta_0)$.
Next, we can perform the standard in-in formalism to compute the one-loop bispectrum. 
The bulk-to-boundary propagators of $\phi$ are given by 
\begin{equation}
    K_+(k,\eta)=\phi_k(\eta_0)\phi_k^*(\eta)~,~~~~K_-(k,\eta)=K_+^*(k,\eta)~,~~~~{\rm with} ~~\phi_k(\eta)=\frac{H}{\sqrt{2k^3}}(1+ik\eta)e^{-ik\eta}
\end{equation}  
We also need the bulk-to-bulk propagators of $\sigma$: 
\begin{align}
& G_{++}(k,\eta, \eta') = \sigma_k(\eta) \sigma^*_k(\eta') \Theta(\eta - \eta') +
\sigma^*_k(\eta) \sigma_k(\eta') \Theta(\eta' - \eta) ,~~~~ G_{+-}(k,\eta, \eta')=\sigma_k^*(\eta)\sigma_k(\eta')  \\
& G_{--}(k,\eta, \eta')=G^*_{++}(k,\eta, \eta'), ~~~~ G_{-+}(k,\eta, \eta')=G^*_{+-}(k,\eta, \eta') .
\end{align}
with the mode function $\sigma_k(\eta)=({H\sqrt{\pi}}/{2}) e^{i\pi/4+i\pi\nu/2}
 (-\eta)^{3/2}H^{(1)}_{\nu}(-k\eta) $.
Then the Feynman rules lead to the following three-point function
    \begin{align}\label{integral}
 \langle \phi_{{\bf k}_1}\phi_{{\bf k}_2}\phi_{{\bf k}_3} \rangle' & = \frac{1}{8\Lambda^3}\sum_{a_1,a_2,a_3=\pm}\int_{-\infty}^{\eta_0}\prod_{i=1}^3\frac{\mathrm{d\eta_i}}{\left(-H\eta_i\right)^2}
    \Big[a_i\partial_{\eta_i}K_{a_i}(k_1,\eta_1) \delta(\eta_i-\eta_0)\Big] \int\frac{d^3\mathbf{q}}{(2 \pi)^3}\prod_{i<j} G_{a_ia_j}(q_{ij};\eta_i,\eta_j) \nonumber\\
    &  =  \frac{1}{8\Lambda^3}\int\frac{d^3\mathbf{q}}{(2\pi)^3}|\sigma_q(\eta_0)|^2|\sigma_{|\mathbf{q}+\mathbf{k}_2|}(\eta_0)|^2|\sigma_{|\mathbf{q}-\mathbf{k}_3|}(\eta_0)|^2,
    \end{align}
The time integrals become trivial due to the presence of the Dirac delta functions in the boundary-term interaction. 
Consequently, we are left with the momentum integration only. By using the late-time limit
\begin{equation}
    \lim_{\eta_0\rightarrow 0 }\sigma_k(\eta_0) = \frac{H}{2\sqrt{\pi}}\Gamma(\nu)\left(-\eta_0\right)^{3/2}\left(-\frac{k\eta_0}{2}\right)^{-\nu},
\end{equation}
we reproduce the triangle-loop integral in \cref{bispectrumfieldredefi}.

\end{widetext}

\bibliographystyle{apsrev4-1}
\bibliography{ref}

\end{document}